\documentclass{emulateapj}
\usepackage{epsfig,natbib}
\usepackage{amsmath,amsfonts,amssymb}
\usepackage{mathptmx}
\usepackage{apjfonts}
\usepackage{color}
\usepackage{graphicx}
\newcommand{\Msun}{\>{\rm M_{\odot}}}

\newcounter{species}

\begin{document}

\shorttitle{Disruption of a Red Giant by a SMBH}
\title{Disruption of a Red Giant Star by a Supermassive Black Hole and the Case of PS1-10jh}

\author{Tamara Bogdanovi\'c\altaffilmark{1,3,4}, Roseanne M. Cheng\altaffilmark{1,3} 
and Pau Amaro-Seoane\altaffilmark{2,3}}

\altaffiltext{1}{Center for Relativistic Astrophysics, School of Physics, 
Georgia Tech, Atlanta, GA 30332}
\altaffiltext{2}{Max Planck Institut f\"ur Gravitationsphysik 
(Albert-Einstein-Institut), D-14476 Potsdam, Germany}  
\email{tamarab@gatech.edu, rcheng@gatech.edu, Pau.Amaro-Seoane@aei.mpg.de}
\altaffiltext{3}{visitor at the Kavli Institute for Theoretical Physics, University of California, Santa Barbara, CA}
\altaffiltext{4}{Alfred P. Sloan Research Fellow}

\begin{abstract}

The development of a new generation of theoretical models for tidal
disruptions is timely, as increasingly diverse events are being
captured in surveys of the transient sky. Recently, Gezari et al.
reported a discovery of a new class of tidal disruption events: the
disruption of a helium-rich stellar core, thought to be a remnant of a
red giant (RG) star. Motivated by this discovery and in anticipation
of others, we consider tidal interaction of an RG star with a
supermassive black hole (SMBH) which leads to the stripping of the
stellar envelope and subsequent inspiral of the compact core toward
the black hole. Once the stellar envelope is removed the inspiral of
the core is driven by tidal heating as well as the emission of
gravitational radiation until the core either falls into the SMBH or
is tidally disrupted. In the case of tidal disruption candidate
PS1-10jh we find that there is a set of orbital solutions at high
eccentricities in which the tidally stripped hydrogen envelope is
accreted by the SMBH before the helium core is disrupted. This places
the RG core in a portion of parameter space where strong tidal heating
can lift the degeneracy of the compact remnant and disrupt it before
it reaches the tidal radius. We consider how this sequence of events
explains the puzzling absence of the hydrogen emission lines from the
spectrum of PS1-10jh and gives rise to its other observational
features.

\end{abstract}

\keywords{accretion -- black hole physics --  galaxies: nuclei}

\section{Introduction}

A star whose orbit takes it on a close approach to a supermassive
black hole (SMBH) will be torn apart when the SMBH's tidal force is
comparable to the self-gravity of the star \citep{rees88}. The
subsequent accretion of stellar debris by the SMBH generates a
characteristic flare of UV and soft X-ray radiation, a signature which
for nearly two decades has been used as a strong diagnostic for the
presence of an SMBH in a previously inactive galaxy
\citep[][etc.]{grupe95a,grupe95b,grupe99,renzini95,bade96,kg99,greiner00,donley02,li02,komossa04,gezari06,gezari08,gezari09,velzen11}. For
many of the observed candidates, the appearance of the flare and its
properties can be explained by an encounter of a main sequence star
with the SMBH, which is expected to be the most common disruption
scenario because of the relative abundance of this population of
stars.  More recently however, the improved observational coverage of
the transient sky led to discoveries of new classes of tidal
disruption events.

One such event (PS1-10jh) was reported by \citet{gezari12} as a disruption
of a helium-rich stellar core, thought to be a remnant of a red giant (RG)
star.  PS1-10jh shares some common properties with previously detected 
candidates which supports its identification as a tidal
disruption, but also exhibits differences which may be a key to
understanding the physical processes at play. For example, a
well-sampled UV light curve of PS1-10jh is in agreement with the
canonical theoretical expectation that the luminosity of a tidal
disruption event should decay as $\propto t^{-5/3}$
\citep{rees88,ek89}\footnote{The power-law behavior of tidal
disruption light curves has however been predicted to vary as a
function of time and emission wavelength \citep{lodato09, lr11}.}.
The optical spectrum of the event is characterized by the presence of
relatively broad HeII emission lines, a property which is unusual
given the absence of broad hydrogen emission lines. Both sets of
emission lines are commonly found in the spectra of active galactic
nuclei (AGNs), where hydrogen Balmer lines typically dominate in
strength because of the higher abundance of hydrogen relative to
helium in the interstellar medium. In light of this, \citet{gezari12}
interpreted the optical spectrum of PS1-10jh as the emission from the
disrupted He core of an RG star which lost its hydrogen envelope at
some earlier point in time. Similar interpretation was provided
by \citet{wang11,wang12} who discovered another tidal disruption
candidate with strongly blueshifted HeII emission line and weak
hydrogen lines in their sample of extreme coronal line emitters.

Indeed, a tenuous outer envelope can be tidally stripped from the
progenitor RG star at relatively large distances from the SMBH and
before its compact core is disrupted or swallowed. This implies that
the hydrogen envelope debris in the aftermath of disruption can be
distributed over a wide range of spatial scales (in some cases
comparable to the size of the broad line region (BLR)). Consequently, like
the BLR, it is expected to be characterized by a
range of ionization parameters -- a state which in general does not
preclude emission of hydrogen lines. Hence, one of the intriguing
questions about PS1-10jh is {\it what happened to its hydrogen
envelope}?

In this paper we adopt a premise that tidally disrupted star 
was a red giant and investigate how this assumption constrains the
orbital parameters and structure of the star that fell towards the
black hole. Specifically, we consider scenarios that can result in
early separation and subsequent accretion of the hydrogen envelope of
the star, so that at a later point in time when its helium core is
disrupted, helium dominates the emission line spectrum. It is worth
noting that this is not the only theoretical scenario put forward to
explain the nature of PS1-10jh. \citet{guill13} propose that the
emission-line spectrum of PS1-10jh may be produced in the disruption
of a low mass main sequence star. In this case, the debris from the
disrupted star forms a compact BLR that consists of hydrogen rich gas,
but is highly ionized and effectively behaves as a HII region.
Consequently, in this scenario it emits weak hydrogen emission lines
below the current detection threshold.

An interesting property of RG encounters with SMBHs is that they can
lead to partial disruptions with a surviving compact core.  In
scenarios where the stellar core is deposited deep in the potential
well of the SMBH its dynamics will be affected by the relativistic
precession of the pericenter, precession of the orbital plane due to
the spin of the SMBH, and emission of gravitational waves (GWs). A
stellar compact object that gradually spirals into a SMBH is commonly
referred to as the extreme mass-ratio inspiral (EMRI).  GW
emission from EMRIs will only be detected by a future space-based
gravitational wave interferometer \citep[e.g.,][]{as13a}.
Nevertheless, if relativistic effects leave a visible imprint in the
EM signatures of the disruption event, they can in principle be used
to place independent constraints on the structure of the star and mass
and spin of the SMBH \citep{gomboc05,kesden12a,kesden12b} many years
in advance of such GW observatory.

More generally, RG tidal disruptions are unique among stellar
disruption scenarios because they can in principle serve as a tracer
of a wide range of SMBHs.  This is because the low density RG envelope
can be tidally stripped even by the most massive of black holes which
swallow, rather than disrupt main sequence and compact stars.  Indeed,
RGs are expected to make up a dominant fraction of the stellar diet
for SMBHs with mass $\gtrsim 10^{8}M_\odot$ and about 10\% of it for
black holes below this threshold \citep{macleod12}.  They are also
expected to contribute significantly to the duty cycle of low-level
activity of SMBHs with mass $>10^7M_\odot$ \citep{macleod13}. While
disruptions of main sequence stars by $\sim 10^6 M_\odot$ SMBHs will
continue to dominate the detection rates, a growing observational
sample of tidal disruption events indicates that more RG disruptions
are likely to be observed in the future.

In this work we consider a partial disruption scenario where the
stripping of the hydrogen envelope is followed by the subsequent
inspiral and possible disruption of the compact remnant. In the case
of PS1-10jh we find a set of orbital solutions where tidal interaction
with an SMBH leads to the inspiral and disruption of the helium core
after the accretion of the hydrogen envelope has been completed.  Such
configuration can provide an explanation for the lack of hydrogen
emission signatures in the spectrum of PS1-10jh and places strong
constraints on the dynamical evolution and structure of the core.

The remainder of this paper is organized as follows: in
Section~\ref{S_stripping} we describe the stripping of the red giant
envelope and give the physical properties of the debris in
Section~\ref{S_properties}. We consider the inspiral of the remnant
core into the SMBH in Section~\ref{S_inspiral}, discuss implications
for the disruption candidate PS1-10jh in Section~\ref{S_ps10}, and
conclude in Section
\ref{S_conclusions}.

\section{Tidal stripping of the red giant envelope}\label{S_stripping}

Consider a relatively common, low mass RG star with a zero-age main
sequence mass of $M_* = 1.4\Msun$ and radius $R_* \approx
10^{12}\,{\rm cm}$\footnote{\citet{macleod12} have recently modeled
the dynamics of a red giant of this structure during one partially
disruptive encounter with a supermassive black hole.}.  This class of
RG stars consists of a degenerate helium core, $M_{\rm core} \approx
0.3\Msun$ and $R_{\rm core}\approx 10^9\,{\rm cm}$, and a hydrogen
rich envelope with mass $M_{\rm env} \approx 1.1\Msun$, where the mass
and size of the RG core are consistent with the structure of low-mass
helium white dwarfs \citep{hs61,ab97}. An RG star of this mass spends
${\rm few}\times 10^8$~yr on the RG branch of the Hertzsprung-Russell
diagram and transitions onto the horizontal branch after the ignition
of the helium core in the event known as the helium flash.

The conditions for the tidal disruption of a star are fulfilled at the
{\it tidal radius}, $R_T \approx R_*(M/M_*)^{1/3}$, where $M$ is the
mass of the black hole \citep{rees88}.  The tenuous and spatially
extended stellar envelope renders RG stars particularly vulnerable to
tidal forces, resulting in tidal radii $\sim 100$ times larger than
for main sequence stars,
\begin{equation}
R_T \approx 10^{14}\,{\rm cm}\left(\frac{R_*}{10^{12}\,{\rm cm}}\right)
\left(\frac{M_*}{1\,\Msun}\right)^{-1/3}
M_6^{1/3}.
\label{eq_rt_cm}
\end{equation}
In terms of gravitational radii,
\begin{equation}
R_T \approx 680\,r_g\left(\frac{R_*}{10^{12}\,{\rm cm}}\right)
\left(\frac{M_*}{1\,\Msun}\right)^{-1/3}
M_6^{-2/3},
\label{eq_rt_rg}
\end{equation}
where $M_6=M/10^6\Msun $ and $r_g = GM/c^2 = 1.48\times 10^{11}\,{\rm
cm}\,M_6$.  More specifically, this disruption radius is defined for a
gas cloud whose density is equal to the {\it mean density} of an RG
star. Since the mean density of the RG star is comparable to that of
its hydrogen envelope, Equations \ref{eq_rt_cm} and \ref{eq_rt_rg}
provide a good estimate for the distance from the SMBH where the
stellar envelope will be stripped by the tides. On the other hand, a
white dwarf-like core can survive the tidal forces of a million solar
mass black hole even at separations comparable to the radius of its
event horizon (equal to $2r_g$ for non-rotating black holes).

For a star on an orbit with pericentric distance $R_p$, the strength
of the tidal encounter is characterized by the penetration factor
$\beta = R_T/R_p$, where $\beta =1$ marks the threshold for tidal
disruption and $\beta \gg 1$ characterizes deep encounters
\citep{cart82}.  Similarly, the penetration factor for the RG core is
defined as
\begin{equation}
\beta_{\rm core} = \beta \;\left(\frac{R_{\rm core}} { R_*}\right)\left( \frac{M_*}{M_{\rm core}}\right)^{1/3} .
\label{eq_betac}
\end{equation}
For the core under consideration, $\beta_{\rm core} \approx
\beta/600$.  The core of finite extent experiences the gradient of the
SMBH gravitational potential which at pericenter leads to a spread in
the specific gravitational potential energy across the core's radius
of $\Delta \epsilon_{\rm core} \approx GMR_{\rm core} / R_p^2$
\citep[][etc.]{lacy82,ek89}. Recent calculations however show that 
for disruptive, $\beta > 1$ events the energy spread of the debris
remains frozen at the point of disruption and is a very weak function
of $\beta$ \citep{stone12,grr13}.This implies that the envelope which disrupts
at $\sim R_T$ has energy spread $\Delta \epsilon_{\rm env} \approx
GMR_*/ R_T^2$.  For nearly parabolic orbits the most bound portion of
the envelope debris is characterized by $\epsilon = -\Delta
\epsilon_{\rm env}$ and shortest Keplerian orbital period 
\begin{equation}
P_m\approx 2\times 10^8{\rm s}\,\,
\left(\frac{R_*}{10^{12}{\rm cm}} \right)^{3/2}
\left(\frac{M_*}{1\Msun} \right)^{-1}
M_6^{1/2}.
\label{eq_pm}
\end{equation}

Assuming the distribution of debris mass over energy is approximately
symmetric, $M_*/d\epsilon \approx f M_*/(2\Delta\epsilon)$
\citep{ek89}, it is possible to derive the accretion rate of the
debris bound to the SMBH
\begin{eqnarray}
&&  \dot{M}  = \frac{dM_*}{d\epsilon}\frac{d\epsilon}{dt} 
= \frac{f\,M_*}{3P_{\rm m}} \left(\frac{t}{P_{\rm m}}\right)^{-5/3}\approx \nonumber \\
 & &  0.054\,\Msun\,{\rm yr^{-1}}\,f
\left(\frac{R_*}{10^{12}{\rm cm}} \right)^{-3/2}
\left(\frac{M_*}{1\Msun} \right)^{2}
M_6^{-1/2}
\left(\frac{t}{P_{\rm m}} \right)^{-5/3},
\label{eq_mdot}
\end{eqnarray}
where $f$ is the fraction of the RG stellar envelope that is accreted
relative to the mass of the star.

\citet{macleod12} used 3d hydrodynamic simulations to model the fraction of the red 
giant stellar envelope lost to tidal stripping as a function of the
penetration factor of the encounter. They find that for $\beta=1$
encounters the mass loss for a $1.4\Msun$ RG star amounts to about
$0.4\Msun$ while for moderately deep $\beta\geq 2$ encounters it
asymptotes to $0.7\Msun$. The rest of the stellar envelope remains
gravitationally bound to the core. For nearly parabolic orbits
approximately half of the stripped envelope is launched toward the
black hole while a comparable fraction is unbound and ejected from the
system. It follows that after the most tightly bound debris completes
its first orbit, the half that is bound to the SMBH will begin forming
an accretion disk of mass $\sim0.2-0.35\Msun$.

In this work we focus on RG stars on nearly radial initial orbits 
with eccentricities close to unity. This choice is motivated by the
prediction of the loss cone theory that most of the stars on
disruptive orbits originate from the sphere of gravitational influence
of the SMBH or beyond \citep{mt99,wm04},
\begin{equation}
r_h = \frac{GM}{\sigma^2} = 1.3\times 10^{18}\,{\rm cm}
\left(\frac{\sigma}{100{\rm km\,s^{-1}}} \right)^{-2}
M_6
\label{eq_rh}
\end{equation}
where $\sigma$ is the stellar velocity dispersion. In absence 
of any other effects which could modify stellar orbits this implies
that tidal stripping of the RG envelopes is important for stars with
orbital eccentricities $1-e \lesssim R_T /r_h \sim 10^{-4}$ (see
however Section~\ref{S_dynamics} for discussion of dynamical processes
that are expected to affect RG orbits).

Note that if the RG star is initially placed on a less eccentric orbit
around the black hole, more than $\sim 50\%$ of its debris ends up
bound to the SMBH, placing an upper limit on the mass of the accretion
disk as $0.7\Msun$. The disruption of stars on lower eccentricity
orbits will also result in the return of debris to pericenter on time
scales shorter than given by Equation~\ref{eq_pm} and consequently
peak accretion rates larger than shown in Equation~\ref{eq_mdot}
\citep{hayasaki13,dai13}.

\section{Emission properties of the debris}\label{S_properties}

It is worth noting that the envelope loss described above takes place
at the point where the orbital radius of the star is comparable to
$R_T$, which for $\beta>1$ encounters occurs before the pericentric
passage. As the RG remnant (i.e., stellar core plus the portion of the
envelope bound to it) continues to approach pericenter, its self
gravity continues to diminish relative to the tidal force of the black
hole until the point of pericentric passage after which the trend
reverses \citep{cart82,stone12}. The consequence of this
reversal, combined with the rotation of the spun up core, is that the
lobes which are nearly unbound from the remnant fold
around and crash down on the core (R.M. Cheng et al. 2014, in preparation).
This drives strong shocks in the outer layers of the remnant. In order
to estimate an upper limit on the thermal energy produced by shocks
and deposited onto the star during its pericentric passage, we assume
that it is comparable to the binding energy of the envelope that
crashes back onto the RG remnant. From results of \citet{macleod12},
the fraction of the total envelope mass $\eta M_{\rm env}$ that
contracts back to the star is about $\eta = 0.1$ and at most 0.2 for
deeper encounters.

If all of this material is shock-heated the maximum amount
of thermal energy produced can be estimated as $E_{\rm th} \approx
\eta G M_* M_{\rm env}/ R_*$.  Since the envelope has a large optical
depth due to Thomson scattering, $\tau_T \approx 3\eta \,M_{\rm env}
\,\sigma_T/ (4\pi m_p R_*^2) \approx 2\times 10^7$, in this scenario
the thermal energy is radiated gradually over the diffusion time scale
$t_{\rm diff}\sim
\tau_T R_*/c \approx 7\times 10^8\,{\rm s}$. The resulting luminosity
of the flare is then
\begin{equation}
L_{\rm th} \sim \frac{E_{\rm th}}{t_{\rm diff}} \approx 
4\times 10^{37}{\rm erg\,s^{-1}}
\left(\frac{M_*}{1\Msun} \right) \;\;.
\label{eq_Lb}
\end{equation}
A characteristic black body temperature of the shocked star 
and emitted radiation can be estimated from $\sigma T^4 = L_{\rm
th}/4\pi R_*^2$ and is about $kT\sim$1~eV for the RG properties
considered here. Therefore, in the case of the RG tidal disruption,
the shock from the fallback of nearly-unbound material from the tidal
tails leads to an initial optical flare, softer than $\sim$1~keV flare
expected from a completely disrupted main sequence star and
considerably less energetic than expected from the full disruption of
a white dwarf by an intermediate mass black hole \citep{kobayashi04}.
This emission arises in advance of the flare powered by the
accretion of the debris onto the SMBH. In the case of the RG star
considered here, it lasts for about 20~yr and thus, cannot really be
considered a "flare" as long as the slow diffusion processes determine
the time scale for emission of the thermal energy. In reality, not all
of the thermal energy produced in the fallback would be radiated and
some of it would be deposited into the remnant, so the estimated
luminosity in Equation~\ref{eq_Lb} should be considered as an upper
limit.

Additional emission arises from the envelope debris which falls back
toward the SMBH and starts forming an accretion disk. The debris
bound to the hole, initially distributed over a range of highly
eccentric orbits, will circularize once the orbits are closed. The
conservation of angular momentum implies that the debris circularizes
at a radius equal to twice its initial disruption radius, which in the
case of the RG envelope is $a\approx 2R_T$.  This simple estimate is
supported by hydrodynamic simulations that follow the early evolution
of the debris disk \citep{hayasaki13}.  Because shocks are localized
to "intersection points" of the debris streams the radiation emitted
in this process encounters a relatively small column of matter and
should escape to infinity without significant reprocessing by the
intervening debris.  We thus assume that the energy released during
the circularization will be promptly emitted as radiation. From
conservation of angular momentum, it is possible to show that the
amount of orbital energy released during circularization is $\Delta
E_{\rm circ} \approx GMM_{\rm disk} (2e - 1) /(4R_T)\approx GMM_{\rm
disk}/(4R_T)$, where $e\rightarrow 1$ is the initial orbital
eccentricity of the debris. The luminosity produced during
circularization is then
\begin{eqnarray}
&& L_{\rm circ} \sim  \frac{\Delta E_{\rm circ}}{P_m} = 
 7\times10^{41}{\rm erg\,s^{-1}}\times \nonumber \\
& & \left(\frac{R_*}{10^{12}{\rm cm}} \right)^{-5/2}
\left(\frac{M_*}{1\Msun} \right)^{4/3}
\left(\frac{M_{\rm disk}}{0.2\Msun} \right)
M_6^{1/6},
\label{eq_Lcirc}
\end{eqnarray}
where $M_{\rm disk}$ is the mass of the debris disk, corresponding to
the amount of the stellar envelope that remains bound to SMBH (see
discussion at the end of Section~\ref{S_stripping}).

We next consider the fraction of luminosity contributed by the
fall-back of debris from the circular orbit with the average radius of
$\sim2R_T$, all the way to the innermost stable circular orbit (ISCO)
of the SMBH. Note that in reality circularization and inspiral of the
debris play out concurrently rather than in sequence but we consider
the two processes separately in order to estimate their individual
luminosity budgets. Initially, the luminosity of the accretion flow
increases with time as the radiative efficiency of the debris that
inspirals toward ISCO increases,
\begin{eqnarray}
&&L_{\rm fb} = \epsilon(r) \;\dot{M} c^2 \approx 
2\times 10^{44} {\rm erg\,s^{-1}} f\,\left(\frac{\epsilon}{0.057}\right) \times\nonumber \\
&& \left(\frac{R_*}{10^{12}{\rm cm}} \right)^{-3/2}
\left(\frac{M_*}{1\Msun} \right)^{2}
M_6^{-1/2}
\left(\frac{t}{P_{\rm m}} \right)^{-5/3},
\label{eq_Lfb}
\end{eqnarray}
where $\epsilon = 1 - (r-2)/(r(r-3))^{1/2}$ is the radiative
efficiency for a Schwarzschild black hole and $r$ is the orbital
radius of the debris in units of $r_g$ \citep[for e.g.,][]{fkr02}. For
$r=2R_T$ and $R_T$ given by Equation~\ref{eq_rt_rg}, radiative
efficiency is only $\epsilon \approx 4\times 10^{-4}$ and it gradually
increases to 0.057 as material reaches ISCO, which is located at
$6r_g$ for a non-rotating black hole. Therefore, with the exception of
the early stages of circularization when the low radiative
efficiency results in a lower fall-back luminosity, at most times
$L_{\rm fb} \gg L_{\rm circ} \gg L_b$. According to
Equation~\ref{eq_Lfb}, at ISCO the debris reaches its peak luminosity
which for low mass SMBHs is comparable to their Eddington
luminosity\footnote{Eddington luminosity is defined as $L_{\rm
Edd}\approx 1.3\times10^{44}\,{\rm erg\,s^{-1}}M_6$.}. This implies
that radiative pressure can play an important role in limiting the
luminosity of such systems.

The temperature of the debris radiating at the Eddington limit from
the radius $\sim 2R_T$ is \citep{ulmer99}
\begin{eqnarray}
&& T_{\rm eff} \approx \left(\frac{L_{Edd}}{16\pi R_T^2 \sigma}\right)^{1/4} = \nonumber \\
&&  4.6\times10^4\,{\rm K} \,
\left(\frac{R_*}{10^{12}{\rm cm}} \right)^{-1/2}
\left(\frac{M_*}{1\Msun} \right)^{-1/6}
M_6^{1/12} .
\label{eq_Teff}
\end{eqnarray}
where $\sigma$ is the Stefan-Boltzmann constant. As material finds
itself deeper in the potential well of the black hole, the temperature
of the emitted multi-temperature black-body radiation increases by a
factor of $(2R_T/ 6 r_g)^{1/2}$, or about 15 times for the scenario
considered here. Therefore, the temperature evolution of the radiation
emitted during the fall-back phase can in principle be used to infer
the radiative efficiency of the debris disk at different radii from
the SMBH. In practice however, the emitted radiation is likely to be
reprocessed by the debris material and emerge as radiation of lower
temperature.  This occurs because the debris disk with radius
$\sim2R_T$ has a significant optical depth, $\tau_T \approx M_{\rm
disk} \;\sigma_T / 2 \pi (2R_T)^2 m_p \approx 730 $. Nevertheless, if
the broad emission line profiles (such as FeK$\alpha$ or hydrogen
Balmer lines) are available from spectroscopic observations for this
early phase of disruption before the luminosity peak, their width and
shape can be used to constrain the radius of the inspiralling debris
independently from its temperature.
 
The evolution of optical depth of the debris in the scenario
considered here implies that the debris disk will become transparent
to radiation once it expands by a factor of $\sim 27$ due to the
transport of angular momentum. The conservation of angular momentum
implies that at this point in time the mass of the disk will be $\sim
M_{\rm disk}/\sqrt{27}$, after about $80\%$ of its mass has been
accreted onto the black hole. Assuming the accretion rate given by
Equation~\ref{eq_mdot}, for a $\beta=1$ encounter of a $1.4\Msun$ RG
with a $10^6\Msun$ SMBH and $M_{\rm disk} = 0.2\Msun$ this amount of
debris is accreted only $t_{\rm acc}\sim {\rm few}\times\,P_m$ after
the luminosity peak, at which point the character of the debris disk
emission changes to optically thin. In this simple estimate we only
consider the fall-back accretion rate and direct the reader to a
recent work by \citet{sm13} for a careful treatment of evolution of
transient disks that form in tidal disruptions.

\section{Inspiral of the compact core into a supermassive black hole}\label{S_inspiral}

We now estimate the orbital properties of the surviving stellar
remnant given a range of initial orbits as a function of the
penetration factor and orbital eccentricity. The range of penetration
factors considered here corresponds to the orbital pericenters between
the ISCO (where the core plunges into the SMBH) and $R_T$ (where the
RG star loses its envelope). For Schwarzschild black holes of
different masses this range corresponds to
\begin{eqnarray}
10^6 M_\odot : R_p & \approx & 9\times 10^{11} - 10^{14}\,{\rm cm} \nonumber \\
10^7 M_\odot : R_p & \approx & 9\times 10^{12} - 2\times10^{14}\,{\rm cm} \label{eq_rp_range} \\
10^8 M_\odot : R_p & \approx & 9\times10^{13} - 5\times10^{14}\,{\rm cm} \nonumber.
\end{eqnarray}
The mechanisms that potentially determine the orbital evolution of the
RG remnant in this phase are: (a) the interactions with the stars in
the nuclear star cluster (b) tidal interaction with the SMBH, (c)
emission of gravitational waves, and (d) gravitational and
hydrodynamical interaction with the disk formed from the envelope
debris.

\subsection{Orbital Dynamics of the Core in the Galactic Nucleus}\label{S_dynamics}

Stars on highly eccentric orbits can venture far from the central SMBH
and experience a gravitational influence of other stars in the nuclear
cluster at apocenter, where gravitational influence of the SMBH is at
the minimum. Their orbital energy and angular momentum changes as a
consequence of these interactions and therefore, it is necessary for
such spatially extended orbits to consider the contribution of the
cluster stars to the gravitational potential of the central SMBHs.

In crowded stellar environments, in the centers of nuclear star 
clusters, the initial energy and angular momentum of an RG star can 
also be altered by physical collisions with other stars
\citep{fr76}. Collisions between RGs and stellar mass black holes have
indeed been suggested as a plausible mechanism that can deplete the
low-mass giants and asymptotic giant branch stars in the mass range
$1-3\Msun$ within the 0.4~pc radius in the Galactic Center
\citep{dale09,davies11}. These studies indicate that the probability
that an RG star will encounter a stellar mass black hole (or with a
lesser probability a main sequence star) at least once during its
lifetime is substantial. Specifically, as the stellar black hole
passes close to the RG, the core is deflected from its original orbit
retaining some fraction of the envelope. A scattering (and even more
so a physical collision) with a massive perturber can however lower
the initial eccentricity of the RG orbit and place it on a bound orbit
around the SMBH, even if it initially originated beyond the sphere of
gravitational influence of the SMBH \citep{perets09}. We thus
conjecture that the RG remnant underwent such an interaction and
consider its end as a starting point of our investigation, given that
the RG remnant has been deflected onto a new orbit with the pericenter
in the range given by Equation~\ref{eq_rp_range} and an arbitrary
apocenter somewhere within $r_h$ (Equation~\ref{eq_rh}).

We borrow from the investigations of EMRIs to elucidate the necessary
conditions for an RG remnant to spiral in close to the SMBH where it
can be disrupted or become a source of GWs.  EMRI studies indicate
that the evolution of compact objects with the range of pericenters
shown in Equation~\ref{eq_rp_range} is dominated by a combination of
relativistic precession effects, GW energy loss, but also vectorial
resonant relaxation and two body relaxation \citep{as13b}.  Recent
work on stellar dynamics has shown that the formation rate of EMRIs
sensitively depends on their eccentricity. For example, the rates of
the EMRIs with semi-major axes $a_{\rm sb} \sim 10^{15}{\rm
cm}\,(1-e^2)^{-1/3}$ can be strongly suppressed by the presence of
what is usually referred to as the Schwarzschild barrier
\citep{merritt11,brem12}. This barrier is a consequence of 
relativistic precession, which becomes pronounced as the compact
object finds itself sufficiently deep in the potential well of the
SMBH. The precession effectively blocks the gravitational influence of
cluster stars that would otherwise build up over many approximately
coherent orbits \citep{rt96} and drive compact objects into the GW
dominated regime. In the context of the RG remnant the relevant
question is: {\it can the core continue to spiral in toward the SMBH
given the presence of the Schwarzschild barrier?} Along similar lines,
vectorial resonant relaxation has been found to change the orientation
of the stellar orbit relative to the SMBH spin, causing the compact
object that was initially outside of the ISCO at one instance of time
to plunge into the black hole in the next, thus limiting any tidal
disruption or GW signal.

It has been shown that stellar objects with $a > a_{\rm sb}$, 
on highly eccentric orbits can evolve into EMRIs or tidal disruptions
without danger of premature plunge into the SMBH or blockage by the
Schwarzschild barrier \citep{merritt11,brem12,as13b}. We use this
criterion to further constrain the orbit of the RG remnant. For these
objects the two body relaxation and emission of gravitational
radiation remain the main drivers of orbital evolution. Which
mechanism dominates is determined by the density and distribution of
stars in the nuclear star cluster, the mass of the SMBH, and the
initial orbit of the RG core. 

The two body relaxation is a diffusive (random walk) process 
resulting from small-angle gravitational scattering following the
encounter of two stars \citep{chandra42}, that can change both the
orbital energy and angular momentum of the scattered star. The
relaxation in angular momentum occurs on time scale that is $(1-e^2)$
times shorter than that for the energy, implying that stars on very
eccentric orbits can significantly change their eccentricity while
maintaining approximately the same semi-major axis \citep{amk12}. In
the case of an RG remnant which is undergoing two body scattering close
to the apocenter of its orbit this would cause the pericentric
distance to change. Since we are interested in following the cores
that approximately maintain their pericentric radius over multiple
passages by the SMBH, we need to determine {\it which orbits are
likely to be affected by two body relaxation as opposed to the
emission of GWs}.

We use Equation~6 from \citet{amk12} and estimate the critical
pericentric distance for return orbits at which the time scale for two
body relaxation in angular momentum is equal to the time scale for
emission of gravitational radiation to shrink and circularize the
orbit,
\begin{equation}
R_p^{\rm crit} \sim 10^{12}{\rm cm} \left(\frac{a}{10^{18}{\rm cm}}\right)^{-1/2}\,M_6^{5/4}
\label{eq_crit_rp}
\end{equation}
for a cluster with the assumed stellar mass density profile $\rho(r)
\propto r^{-7/4}$ \citep{bw76}.
The orbits with pericenters below this critical value will evolve
predominantly due to the emission of gravitational waves and those
above it will evolve due to the two body interactions. Note that
a large portion of the parameter space outlined in
Equation~\ref{eq_rp_range} (but not all of it) resides within the GW
dominated regime as long as $a<10^{18}{\rm cm}$.  Hence, there is an
allowed region of the orbital parameter space where the RG remnant
will continue to spiral into the SMBH due to the emission of GWs, such
that $10^{15} < a < 10^{18}{\rm cm}$ and pericenter is in the range
given by Equation~\ref{eq_rp_range}. In the rest of the paper we focus
on RG remnants which find themselves on such orbits.
  
\subsection{Evolution Due to the Stripping of the Stellar Envelope}\label{S_envelope}

The 3d relativistic hydrodynamic simulations of single encounter between
a white dwarf and an intermediate mass black hole by \citet{cheng13}
indicate that during partial disruption most of the tidal energy is
deposited into the surface layers of the star by shock heating.
Radiation can escape the surface layers of the star easier than its
denser central parts, which implies that shocks during repeated
pericentric passages should cause a characteristic brightening of the
RG remnant. As a result, the luminosity, temperature, and time scale
for radiation to diffuse out of the surface layers of the star would
be similar to those estimated for the fallback of nearly-unbound
material from the tidal tails in Section~\ref{S_properties}. In cases
when the estimated radiation diffusion time scale ($t_{\rm diff}
\approx 7\times 10^8{\rm s}$) is longer than the orbital period of a
returning orbit, the rate of tidal heating will be higher than the
rate of radiative cooling leading to the net heating of a star. The
orbital period of a Keplerian orbit, $2\pi \sqrt{a^3/GM}$, is
\begin{equation}
P_a = 5\times10^8{\rm s}\; \beta^{-3/2}
\left(\frac{1-e}{0.01}\right)^{-3/2} 
\left(\frac{R_*}{10^{12}{\rm cm}}\right)^{3/2} 
\left(\frac{M_*}{1\Msun}\right)^{-1/2}\;,
\label{eq_period}
\end{equation}
and hence $P_a < t_{\rm diff}$ for $e \lesssim 0.99$ orbits around
$10^6\Msun$ black holes. The implication of this effect is the
increase in luminosity of the stellar remnant {\it as well as} the
expansion and degeneracy lifting in the outer layers.  Because the
luminosity of the shocked outer layers of perpetually heated star can
become comparable or larger than the stellar Eddington luminosity
after only a few orbits (see Equation~\ref{eq_Lb}), the combined
effect of {\it radiative pressure} and {\it expansion} will drive the
increased mass-loss of the remainder of the stellar envelope.
These two regimes have been previously described in some detail 
by \citet{alexander03} who studied a class of tidally heated stars
called {\it squeezars}. Thus, a majority of tidally stripped RG
remnants may lose all of their bound hydrogen envelope after several
orbital passages by the SMBH. A more precise time scale on which this
happens depends on a complex interplay of radiative processes,
hydrodynamics, and orbital dynamics and can only be captured though
simulations.

We carry out a simple estimate to determine the relative importance of
the disruption of the envelope versus the GW emission for the
evolution of the remnant's orbit in the case where the stellar
envelope is removed gradually.  We assume that the envelope is
stripped over $N_s$ orbits, and write the necessary energy budget as
\begin{equation}
\Delta E_s \sim \frac{1}{N_s}\,\frac{\eta_{\rm env} GM_{\rm core}\,M_*}{R_{\rm *}},
\label{eq_dEs}
\end{equation}
where $\eta_{\rm env} = M_{\rm env}/M_*$. We evaluate the energy loss per orbit due 
to the emission of gravitational waves as
\begin{equation}
\Delta E_{\rm gw} = -\frac{64\pi}{5} \frac{G^{7/2}\,M^2\,M_{\rm core}^2\, 
(M + M_{\rm core})^{1/2}}{c^5\,a^{7/2}} f_1(e),
\label{eq_egw1}
\end{equation}
where $f_1(e) = (1 + 73e^2/24 + 37e^4/96)/(1-e^2)^{7/2}$
\citep{peters64}. In the limit when $e\rightarrow1$ this expression
becomes
\begin{equation}
\lim_{e\rightarrow1}\Delta E_{\rm gw} = -\frac{85 \pi}{12 \sqrt{2}}  
\frac{G^{7/2}\,M^2\,M_{\rm core}^2\, (M + M_{\rm core})^{1/2}}{c^5\,R_p^{7/2}} \;.
\label{eq_egw2}
\end{equation}

For most orbital parameters considered here the energy loss due to the
emission of gravitational radiation is well approximated by the
solution based on semi-Keplerian orbits as described in
\citet{peters64}.  However, for cores that make very close pericentric
passages by the SMBH ($R_p\lesssim 10\,r_g$) this approximation ceases
to be accurate. \cite{Martell2004} compared approximations equivalent
to the approach by \citet{peters64} and more accurate prescriptions
and found that the difference in the GW energy loss amounts to a
factor between 0.8 and 1.3. This level of approximation is appropriate
for our study and we therefore use the Peters approximation.

Combining Equations~\ref{eq_dEs} and \ref{eq_egw2} in the limit of
$e\rightarrow1$ we obtain
\begin{eqnarray}
&&\frac{\Delta E_s}{\Delta E_{\rm gw}} \sim 3\times 10^4 \,\eta_{\rm env} \,\beta^{-7/2}
\left(\frac{N_s}{10^4} \right)^{-1}
\left(\frac{R_*}{10^{12}{\rm cm}} \right)^{5/2} \times \nonumber \\
&&\left(\frac{M_*}{1\Msun} \right)^{-1/6}
\left(\frac{M_{\rm core}}{0.3\Msun} \right)^{-1}
M_6^{-4/3}.
\label{eq_Eratio}
\end{eqnarray}

Therefore, hydrodynamic effects related to the stripping of the
envelope are the dominant driver of the orbital evolution in a
majority of scenarios. It is worth noting that if most of the tidal
energy (Equation~\ref{eq_dEs}) is absorbed by the core, than repeated
tidal interactions can in principle supply a significant fraction of
energy necessary to lift the degeneracy of the core. This follows from
the energy budget necessary to generate thermal pressure comparable to
the degenerate pressure of the core, which can be estimated from the
virial theorem of a white dwarf in equilibrium. We consider the tidal
heating of the core in more detail in the next section.

It is worth noting that there is an additional mechanism that 
may drive an increased mass loss from the surface of the star. We 
do not consider the Roche lobe overflow in this work but it has been 
shown by  \citet{macleod13} that RG stars on eccentric orbits continue 
to lose mass from their envelope at each pericentric passage as they 
overflow the Roche lobe for a short time. For an RG on a relatively
stable orbit the mass loss eventually ceases due to the stabilizing 
gravitational influence of the compact core.

Recent simulations also indicate that tidal disruption is an
inherently asymmetric process and that unequal fractions of the debris
are launched toward the SMBH and unbound from the system
\citep{cheng13,manukian13,kyutoku13}.  This asymmetry is
more pronounced (in geometric sense) for the larger mass ratios
($q\gtrsim10^{-4}$) and arises from the fact that more material escapes
from the surface of the star facing the SMBH, relative to the surface
that faces away \citep[see Figures~6 and 7 in][]{cheng13}.
Since unequal portions of the debris carry unequal amounts of linear
momentum, the conservation of linear momentum then requires that the
remnant core is deflected from its initial orbit.  We neglect this
effect which stems from the hydrodynamic interaction of the star with
the tides of the SMBH and assume that the core continues to orbit
along the geodesic previously occupied by the center of mass of the RG
star.

\subsection{Tidal Heating of the Core and Emission of Gravitational Waves}\label{S_theating}

In this section we consider the effect of tidal interaction on the
helium core. The SMBH tides drive oscillations which if dissipated
within the core lead to a gradual circularization of its orbit. It can
be shown that the amount of energy available from circularization is
much larger than the gravitational energy of the star
\citep{ivanov07}. Namely, the energy released from
circularization of an orbit with initial pericentric radius $R_p$ and
eccentricity $e\rightarrow1$ is $\Delta E_{\rm circ} \approx G M
M_{\rm core}/4R_p$. Comparing that to the gravitational energy of the
core, $E_{\rm core} \approx GM_{\rm core}^2/R_{\rm core}$, yields
\begin{eqnarray}
\frac{\Delta E_{\rm circ}}{E_{\rm core}} 
&&\nonumber
\approx \frac{1}{4}\beta_{\rm core}\left(\frac{M}{M_{\rm core}}\right)^{2/3}
\\
&&
\approx 6\times 10^3 \beta_{\rm core}\left(\frac{M_{\rm core}}{0.3\Msun} \right)^{-2/3}\,M_6^{2/3} \;.
\label{eq_ecirc2}
\end{eqnarray}
Even if a small fraction of this energy is deposited into the remnant
it is in principle sufficient to lift the degeneracy of the core and
possibly dissolve it before the core reaches the tidal radius or
plunges into the SMBH \citep{ivanov07}. Recall however that at
the same time the emission of gravitational radiation is concurrently
acting to decrease the orbital eccentricity of the core. If tidal
heating is the faster of the two processes, the remnant will be
destroyed before the orbit is circularized. If however the GW emission
is more efficient, the remnant will find itself on a quasi-circular
orbit, gradually inspiralling into the SMBH and possibly forming an
EMRI \citep[see for example][who studied mass transfer between an
inspiralling white dwarf and a low-mass SMBH in this
regime]{zalamea10}.

To estimate the time scale on which the tides are depositing energy
onto the star we rely on the linear theory of tidal interactions
developed for stars and planets on eccentric orbits
\citep{pt77,leconte10}. The linear theory applies to weak encounters,
rather than disruptive ones when description of the tidal interaction
as a low level perturbation ceases to be accurate.  Consequently, the
linear theory can be used to estimate the energy deposited onto a
compact core, for which $\beta_{\rm core} \ll 1$, but is not
applicable to the stellar envelope that is in the $\beta \gtrsim1$
regime.  In this calculation we therefore consider the tidal heating
of the compact core only and do not account for the effects of the
envelope (which are discussed in Section~\ref{S_envelope} as an order
of magnitude estimate). This is a conservative assumption which leads
to a slower orbital evolution of the core since in absence of the
envelope $\dot{E}_{\rm tid}$ is only a lower limit on the rate of
energy deposition into the remnant. Similar to \citet{amk12}, we adopt
the expression from \citet{leconte10} for the rate at which the
orbital energy is lost to the excitation of oscillatory modes within
the star
\begin{equation}
\dot{E}_{\rm tid} = 2K_p\left [ N_a(e) - \frac{N^2(e)}{\Omega(e)} \right] ,
\end{equation}
where the expression in the brackets only depends on eccentricity and
can be simplified to $7e^2/2$ for $e<1$. This is an adequate
approximation for the orbital evolution of the helium core, which
starts on a highly eccentric orbit and is gradually circularized by
tidal heating and GWs.  Using their expression for $K_p$
\begin{equation}
K_p \approx \frac{9}{4}Q^{-1}\,
\left(\frac{GM_{\rm core}^2}{R_{\rm core}}\right)
\left(\frac{M}{M_{\rm core}}\right)^2
\left(\frac{R_{\rm core}}{a}\right)^6
\left(\frac{GM}{a^3}\right)^{1/2}.
\end{equation}
Modifying further to relate $a$ to the orbital pericenter and orbital
period we obtain
\begin{equation}
\dot{E}_{\rm tid} \approx \frac{63\pi}{2} Q^{-1}\,e^2\,\beta_{\rm core}^6 
\left(\frac{GM_{\rm core}^2}{R_{\rm core}}\right)\,P_a^{-1},
\label{eq_etid_dot}
\end{equation}
where $Q$ is the quality factor, a parameterization of the coupling of
the tidal field to the stellar core. The magnitude of $Q$ is quite
uncertain and commonly assumed in the literature to be in the range
$\sim10^5-10^6$. However, a recent work by \citet{pm12} indicates that
the quality factor of helium white dwarfs may be even higher (they
find $Q\gtrsim10^7$) but these types of measurements are complicated
by the fact that the value of $Q$ can only be determined as a lower
limit and is coupled to other (uncertain) parameters of the system.

We calculate the characteristic time for circularization of the RG
core orbit assuming that tidal dissipation within the remnant is
efficient and obtain
\begin{equation}
\tau_{\rm tid} = \frac{\Delta E_{\rm circ}}{\dot{E}_{\rm tid}} = 
\frac{1}{126\pi} \frac{Q}{\beta_{\rm core}^6\, e^2}
\left(\frac{M}{M_{\rm core}} \right)^{2/3}P_a.
\label{eq_tau_tid}
\end{equation}
We then calculate the amount of time it takes to circularize the orbit
due to the emission of GWs \citep{peters64}
\begin{equation}
\tau_{\rm gw} = \frac{e}{\dot{e}} = 
\frac{15}{608\pi}\,f_2(e)\,\beta_{\rm core}^{-5/2}
\left(\frac{R_{\rm core}}{r_g}\right)^{5/2}
\left(\frac{M}{M_{\rm core}}\right)^{11/6}P_a,
\label{eq_tau_gw}
\end{equation}
where $f_2(e) = (1+121e^2/304)^{-1}$. Comparing the two time scales
yields
\begin{equation}
\frac{\tau_{\rm tid}}{\tau_{\rm gw}}\approx
2\times 10^3 \,\frac{\beta_{\rm core}^{-7/2}}{e^2 f_2(e)}\, Q_6 
\left(\frac{R_{\rm core}}{10^{9}{\rm cm}} \right)^{-5/2}
\left(\frac{M_{\rm core}}{0.3\Msun} \right)^{7/6} M_6^{4/3},
\label{eq_tau_compare}
\end{equation}
\begin{figure}[t]
\center{ 
\includegraphics[width=0.5\textwidth]{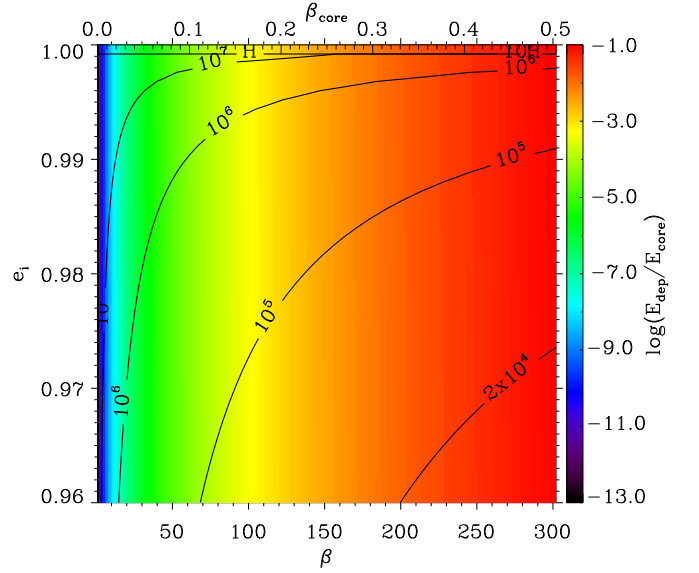} 
\caption{Color illustrates the maximum tidal energy deposited 
into the inspiralling stellar core (in units of its
binding energy) as a function of the initial orbital eccentricity
($e_i$) and strength of the tidal encounter ($\beta$). The top axis
shows the corresponding penetration factor for the core alone. Black
solid lines mark the orbital period of the core in seconds. Orbits in
the top part of the figure have periods longer than the Hubble time,
as marked by the letter "H". }
\label{fig_core}}
\end{figure}

where $10^6Q_6 =Q$. Hence, the gravitational radiation will
always circularize the core's orbit on a time scale much shorter than
that for the tidal circularization. This implies that only a fraction
of the energy available from orbital circularization will be deposited
into the remnant. How big that fraction is in terms of the remnant's
binding energy depends on the properties of the core, its orbit, and
the mass of the black hole. Equations~\ref{eq_ecirc2} and
\ref{eq_tau_compare} for example indicate that close to disruption
($\beta_{\rm core}\sim1$), the core can in principle absorb $\sim
6\times 10^{3} / 2\times 10^3 \sim$ few times its binding energy by
the time its orbit is circularized due to the emission of
GWs. However, the compact core considered here reaches the maximum
penetration factor, $\beta_{\rm core}^{\rm max}\approx 0.5 <1$ at the
event horizon of a $10^6\Msun$ SMBH, where the amount of energy
deposited into it is at maximum $\sim 0.08 E_{\rm core}$.

This is illustrated in Figure~\ref{fig_core} where the color bar 
shows the maximum tidal energy deposited into the inspiralling stellar
core as a function of the initial orbital eccentricity ($e_i$) and
strength of the tidal encounter ($\beta$ and $\beta_{\rm core}$).
This energy will be stored within the core because it is deposited on
the time scale shorter than that on which it can be radiated (i.e.,
the Kelvin-Helmholtz time scale).  Since $\tau_{\rm KH} = E/L$
(where $E$ is the thermal energy of the white dwarf-like core) the
more luminous cores tend to cool faster. Indeed, theoretical
calculations of white dwarf cooling sequences show that a 0.3$\Msun$
white dwarf with luminosity between $0.3 > L > 10^{-4}L_{\odot}$ cools
on a timescale $10^8 < \tau_{\rm KH} < 10^{10}$~yr, respectively
\citep{driebe98,nelemans01}.  For the cores with a representative
cooling time scale of $\tau_{\rm KH}\sim 10^9\,{\rm yr} \gg \tau_{\rm
gw} $ the evolution may proceed similar to {\it squeezars}, tidally
powered stars that find themselves on non-disruptive orbits
\citep{alexander03}.

\subsection{Interaction of the Core with the Envelope Debris Disk}

We now consider the interaction of the RG remnant on a return orbit
with the accretion disk that forms from the tidally stripped hydrogen
stellar envelope.  By the time the core returns, the inner parts of
the debris disk have had time to complete multiple orbits and start
circularizing \citep{ulmer99}. As a consequence, the debris is more
spatially extended than the core's orbit at separation $\sim R_T$ from
the SMBH leading to the intersection of their orbits. Based on
considerations in Section~\ref{S_stripping} we estimate that a core
encounters the debris disk with radius $\sim 2R_T$ and mass in the
range $M_{\rm disk} \approx 0.2-0.7\Msun$, where the lower bound
corresponds to parabolic encounters at the threshold of disruption and
the upper bound corresponds to $\beta>1$ encounters where the RG star
is initially placed on moderately eccentric, bound orbit.

The core can only experience significant orbital energy losses if it
encounters the amount of disk mass comparable to its own. Since the
debris disk is spread to a very low density, the remnant will not be
able to sweep a mass equivalent to its own even after many passages
and thus, the effect of the disk on the orbit of the remnant can be
neglected. This outcome is guaranteed given a relatively short
estimated time scale for accretion of the debris, $t_{\rm acc} \sim
{\rm few}\times P_m$ (see Section~\ref{S_properties}), indicating that
the debris disk will be accreted and dispersed before it can impact
the RG core in any way.

Along similar lines, given a relatively short life time of the
transient disk, it is also unlikely that the core-disk collisions can
give rise to more than a few quasi-periodic luminosity outbursts.

\section{Tidal disruption candidate PS1-10jh}
\label{S_ps10}

\subsection{Constraints on the Properties of the RG Remnant}

We now consider a recently discovered tidal disruption candidate
PS1-10jh described by \citet{gezari12}. PS1-10jh was discovered as an
optical transient in the Pan-STARRS1 Medium Deep Survey and
independently as a near-ultraviolet source within the {\it GALEX} Time
Domain Survey. Based on the modeling of the multi-band photometry and
optical spectra of this object, \citet{gezari12} proposed that the
disrupted object is a tidally stripped helium core of a red giant star
that initially had mass $M_*\gtrsim 1\,\Msun$. They adopt properties
of the core consistent with those measured for an RG stellar remnant
stripped in a stellar binary system \citep{maxted11}, where $M_{\rm
core} \sim 0.23\Msun$ and $R_{\rm core}\sim 0.33\,R_\odot$, and use
the local scaling relations and tidal disruption light curve to
constrain the mass of the SMBH to $M \approx 2.8\times 10^6\Msun$.

Note that the radius of the remnant adopted by \citet{gezari12} is an
order of magnitude larger than that of the "bare" helium core we
consider in previous sections \citep[based on models
of][]{hs61,ab97}. This is because a low-mass red giant remnant created
through stripping and mass transfer in a stellar binary is expected to
retain some fraction of its hydrogen envelope in which hydrogen
burning is maintained in a shell. The luminosity contribution due to
the nuclear burning drives the expansion of the envelope and
degeneracy lifting in the exterior shells of such white dwarf and
results in the radius of the remnant typically a few times larger than
that of a pure helium core \citep{driebe98,driebe99}. 

The choice of the RG remnant structure has important implications for 
the outcome of the tidal interaction. Specifically a bare helium
core considered previously is too compact in size to be disrupted by the SMBH
(Figure~\ref{fig_core}), while the core with the radius $\approx
0.33\,R_\odot$ considered by \citet{gezari12} can be disrupted after
multiple passages (this will be shown in the next section). What is a
more realistic choice for the structure of the remnant is a question
of some subtlety. Even if the RG core initially had a radius of $\sim
10^9{\rm cm}$, given a sudden loss of $30-50\%$ of the RG mass
during its first pericentric passage (see Section~\ref{S_stripping}),
the core will have to adjust to a new equilibrium by expanding beyond
its original radius. We estimate the change in the radius of the core
from the mass-radius relation for the adiabatic evolution of a nested
polytrope, as outlined by \citet{macleod13}.  Depending on the adopted
value of the polytropic index, one finds that the new radius of the
remnant can be up to 5.5 times larger. \footnote{See also
\citet{macleod13} for a discussion of why the response of the remnant
in reality may be different and potentially more dramatic than
indicated by the polytropic model.}  Therefore, we expect that the
radius of the remnant core in the case of PS1-10jh to be larger than
that of the core in an intact RG and for the purpose of the analysis
in this section adopt the mass and radius suggested by
\citet{gezari12}. 

Several observational properties of PS1-10jh offer important clues
about the remnant:
\begin{itemize}
\item The temporal decay of the light curve closely follows 
the canonical prediction $L_{\rm fb}\propto t^{-5/3}$ indicating that
the incoming core must have been on a highly eccentric orbit prior to
disruption such that at least one portion of it ended up being
gravitationally unbound from the system. It has otherwise been shown
that lower eccentricity encounters in which 100\% of the debris is
gravitationally bound to the black hole have the fallback accretion
rate significantly different from $\propto t^{-5/3}$ \citep[][see
however the end of this Section for further discussion of this
point]{hayasaki13,dai13}.

\item An interesting property of PS1-10jh are its broad high-ionization 
He~II lines (with wavelengths $\lambda=4686$\AA~ and
$\lambda=3203$\AA) which are contrasted by the absence of the hydrogen
emissions lines. \citet{gezari12} find that the lack of hydrogen
Balmer lines requires a very low hydrogen mass fraction of $<0.2$,
thus supporting the picture that the disrupted star is a helium
core. The lack of any emission lines from the hydrogen stellar
envelope, which is expected to be several times more massive than the
core in the progenitor RG star, is however puzzling and is the key to
understanding the nature of this event.

\item The characteristic temperature of the emitted radiation from
  PS1-10jh has remained relatively uniform before and after the
  disruption ($T\gtrsim 3\times 10^4$K). The lack of evolution of the
  spectrum indicates that the observed radiation may have been
  reprocessed by the obscuring layer or a shell, enshrouding the inner
  debris disk. Because the luminosity for this system can only be
  determined as a lower limit, $\gtrsim 0.6 L_{\rm Edd}$
  \citep{gezari12}, it follows that both the radiation forces and
  relativistic frame-dragging are viable candidates for a mechanism
  that can distribute the debris across the sky and create
  obscuration.
\end{itemize}

The expectation that the fallback accretion rate differs 
significantly from $\propto t^{-5/3}$ for disruption of stars on
initially bound orbits with moderate eccentricities ($e\lesssim0.99$)
is based on theoretical studies by \citet{dai13} and
\citet{hayasaki13}, who carried out the particle and hydrodynamic
simulations of this scenario, respectively. There is however an inherent 
uncertainty in interpretation of simulated accretion rates in terms of the observed tidal
disruption light curves, since these simulations do not account for
the effects of radiative feedback nor do they model accretion light
curves in a given wavelength band. We acknowledge this uncertainty in
the step where we suggest that an observed optical light curve of
PS1-10jh with behavior close to $L_{\rm fb}\propto t^{-5/3}$ indicates
a similar power-law behavior in the underlying accretion rate.

\subsection{Disruption of the Envelope and Inspiral of the Compact Core}
\label{SS_inspiral}

One possible explanation for the absence of hydrogen emission lines is
that most of the tidally stripped envelope was accreted by the black
hole {\it before} the helium core disruption (see the next
subsection for discussion about other possible explanations). This
scenario could play out if the RG core completed multiple orbits
before being disrupted by the SMBH during which time the envelope was
nearly completely stripped and accreted. We consider the orbital
properties of the incoming RG star which could have given rise to such
a sequence of events and can simultaneously satisfy the described
observational properties. A condition for partial disruption which
must be satisfied in order to have the RG envelope stripped by the
tidal forces and the compact core intact is $\beta\geq 1$ and
$\beta_{\rm core} < 1$. We assume that the stellar core had a
relatively common RG progenitor star with mass and radius $M_* =
1.4\Msun$ and $R_* = 10^{12}\,{\rm cm}$. Note that these choices are
uncertain even though they are physically motivated, as the
evolutionary sequence of the helium remnant cannot be constrained more
precisely from the available data. Based on this adopted structure and
Equation~\ref{eq_betac} we find that the penetration factor of the
compact core is $\beta_{\rm core} \approx \beta / 24$ and thus there
is a range in $\beta$-parameter space for which partial disruption can
be achieved.

In order for most of the hydrogen envelope to be accreted, it must be
placed on {\it bound} eccentric orbits around the SMBH.  It is
possible to estimate from analytic arguments the critical orbital
eccentricity below which the entire envelope is bound.  From the
condition that the orbital energy of the RG envelope (and the star) is
equal to the energy spread in the tidal field of the black hole at the
point on the orbit when the envelope is removed we find
\begin{equation}
 \frac{GM(1-e)}{2R_p} = \frac{GM\,R_{\rm *}}{R_T^2}.
 \label{eq_ecore}
\end{equation}

From this condition, it follows that the initial eccentricity of the
RG stellar orbit must be smaller than $e_{\rm crit}^{\rm env} \approx
1 - (2/\beta) (M_*/M)^{1/3}$ in order for the entire envelope to be
bound to the SMBH \citep{hayasaki13}.  We derive a similar condition
for the first pericentric passage of the RG core,
\begin{equation}
 \frac{GM(1-e)}{2R_p} = \frac{GM\,R_{\rm core}}{R_p^2}.
 \label{eq_ecore}
\end{equation}
Note that in this case the spread in energy reaches maximum at the
pericenter of the orbit, rather than at the tidal radius of the
envelope and that $R_p = R_T/\beta = R_T^{\rm core}/\beta_{\rm
core}$. This implies a critical eccentricity value of $e_{\rm
crit}^{\rm core} \approx 1 - (2 \beta_{\rm core}) (M_{\rm
core}/M)^{1/3}$ and is valid when $\beta_{\rm core} < 1$.

The $\propto t^{-5/3}$ temporal behavior of the disruption light curve
indicates that at least one portion of the tidally disrupted core is
{\it not bound} to the SMBH.  It follows that the eccentricity of the
first orbit of the RG star must satisfy the condition $e_{\rm
crit}^{\rm core} < e_i < e_{\rm crit}^{\rm env}$. This criterion is
satisfied only when $e_{\rm crit}^{\rm core} < e_{\rm crit}^{\rm env}$
leading to
\begin{equation}
\beta \beta_{\rm core} > \left(\frac{M_*}{M_{\rm core}}\right)^{1/3}
\label{eq_criterion}
\end{equation}
and consequently, $ 6 < \beta < 24$, or written in terms of the core
penetration factor, $0.25 <\beta_{\rm core} < 1$. Hence, there is a
portion of the parameter space conducive to disruption of the RG
envelope followed by the disruption of the core upon completion of
some number of orbits by the core.  In such a case, the hydrogen
envelope ends up bound to the SMBH in its entirety, while the debris
created by the subsequent disruption of the helium core is only
partially bound.  It is important to point out that the exact bounds
within which this region lies are approximate, as they are derived
from simple analytic considerations. However, a realization that such
a region must exist for highly eccentric orbits, naturally follows
from the difference in the relative spread in energies of the core and
envelope.

The inspiralling core will be subject to tidal heating and emission of
gravitational radiation, as discussed in
Section~\ref{S_theating}. Taking the ratio of 
Equations~\ref{eq_ecirc2} and \ref{eq_tau_compare} we calculate the
maximum tidal energy that can in principle be deposited into the
remnant before gravitational radiation circularizes its orbit. For the
assumed mass and radius of the helium core disrupted in PS1-10jh and
the inferred mass of the SMBH we find
\begin{equation}
\frac{E_{\rm dep}}{E_{\rm core}}\sim 5\times10^3 \, e^2\, f_2(e)\; 
Q_6^{-1}\beta_{\rm core}^{9/2} \; M_{\rm core}^{-11/6} \; R_{\rm core}^{5/2}\; M^{-2/3}
\label{eq_edep}
\end{equation}
where $M_{\rm core}$, $R_{\rm core}$, and $M$ have been normalized to
the appropriate values for this system. The amount of the deposited
tidal energy is a strong function of $\beta_{\rm core}$. Given that we
are considering the $\beta_{\rm core} < 1$ portion of the parameter space,
where the core gradually inspirals into the SMBH, the ratio of $E_{\rm
  dep}/E_{\rm core}$ will generally be lower than the estimate in
Equation~\ref{eq_edep}.
\begin{figure}[t]
\center{ 
\includegraphics[width=0.5\textwidth]{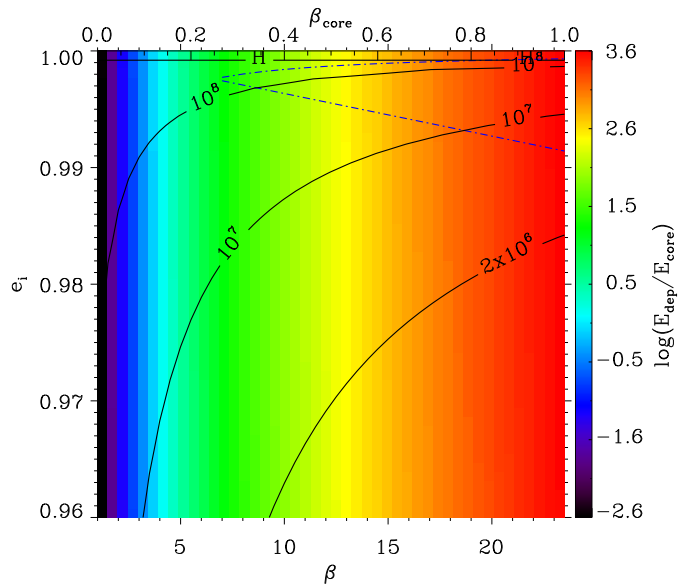} 
\caption{Similar to Figure~\ref{fig_core} except that color here illustrates 
the maximum tidal energy deposited into the inspiralling stellar core 
in PS1-10jh. Blue dash-dotted line marks the region
where RG star is partially disrupted and its hydrogen envelope can be
accreted before the disruption of the compact core.}
\label{fig1}}
\end{figure}

In Figure~\ref{fig1} the color bar shows the maximum tidal energy
deposited into the inspiralling stellar core as a function of the the
initial orbital eccentricity ($e_i$) and strength of the tidal
encounter ($\beta$ and $\beta_{\rm core}$). In this parameter space
$\beta_{\rm core}>0.2$ encounters can in principle deposit the amount
of tidal energy larger than $E_{\rm core}$ prior to orbital
circularization and thus, lift the degeneracy of the core and lead to
its disruption before it reaches the tidal radius (note that this is
different from the outcome shown in Figure~\ref{fig_core}). This can
be accomplished as long as the tidal energy is stored in the core over
many orbits and not efficiently radiated. The initial orbital period
of the core is shown in seconds as a function of $e_i$ and $\beta_{\rm
core}$ with black contour lines. This indicates that the majority of
this parameter space comprises of initial orbital periods $<\tau_{\rm
KH}$. The portion of the parameter space where $e\approx1$ is
characterized by orbital periods longer than the Hubble time is
indicated by the letter "H".

In Figure~\ref{fig1} we also mark the region within which the
criterion for partial disruption described by
Equation~\ref{eq_criterion} is satisfied (blue dash-dotted line).  The
region is confined to a narrow strip of high eccentricities with
values $e>0.991$, characteristic of stellar objects that fall toward
the black hole on a nearly parabolic orbit.  The incoming RG star must
occupy the region below the upper dash-dotted line in order for its
hydrogen envelope to be entirely bound to the SMBH after it is
stripped.  The observed temporal behavior of the light curve in
PS1-10jh on the other hand indicates that at the point of disruption
the helium remnant must still have occupied a very eccentric orbit,
somewhere above the lower dash-dotted line. This region is
characterized by orbits with the period in the range $\sim10^7-10^8$s, 
GW circularization times in the range $\sim10^7-10^9$yr,
and intense tidal interaction with maximum deposited tidal energy in
the range $\sim 10 - 5\times 10^3 E_{\rm core}$. The core occupying
this portion of the parameter space would therefore have had time to
complete many of orbits before the deposited tidal energy
is radiated on a Kelvin-Helmholtz time scale.  It is thus likely that
such a core would have its degeneracy lifted by perpetual episodes of
tidal heating, leading to premature disruption. An early disruption
would also ensure that the core remains on a highly eccentric orbit,
before the GW emission had time to circularize it significantly.

The observed light curve of PS1-10jh also indicates that the 
white dwarf-like core would have to undergo a single disruption event at the
end of the tidal heating period, rather than produce a series of
quasi-periodic accretion events as the core overflows its Roche lobe
at every pericentric passage. The latter scenario has been modeled by
\citet{zalamea10}, who find that the resulting accretion curve
consists of multiple exponentiating flares very distinct from the
$\dot{M}\propto t^{-5/3}$ behavior.  We propose that the core
undergoes a single disruption scenario if the imparted tidal energy
can be redistributed over the volume of the remnant, maintaining the
evolution through a series of quasi-equilibria and avoiding rapid
expansion of the outer layers and stripping.

The time scale on which the tidal energy can be transported from the surface
of the remnant inward is given by its thermal diffusion time scale, 
$\tau_{\rm cond} = R_{\rm core}^2/\chi_{\rm th}\sim 10^6$~yr, where $\chi_{\rm th}$ is the diffusivity constant
\citep{padma01}. The hierarchy of timescales $P_a \ll \tau_{\rm cond}
\ll \tau_{\rm KH}$ implies that tidal interactions establish a
temperature gradient between the surface and the center of the remnant
that last over many orbits but that the tidal energy is diffused
throughout it before the remnant has a chance to cool radiatively. It
is thus plausible in this regime to tidally heat the remnant
gradually, although some amount of tidal stripping from the surface of
the remnant is likely at the rate lower than that found by
\citet{zalamea10}.

\subsection{"Disappearance" of the Hydrogen Envelope} 

In the mean time the stripped stellar envelope forms an accretion disk
with maximum initial mass $M_{\rm disk,i} = M_* - M_{\rm core} \approx
1.2\Msun$, which circularizes and accretes onto the SMBH. The
observations of PS1-10jh indicate that at the moment of disruption of
the helium core the mass fraction of hydrogen must have been more than
five times lower relative to helium and other atomic species thus,
placing the upper limit on the final mass of the hydrogen disk to
$M_{\rm disk,f} < 0.2 M_{\rm core} \approx 0.05\,\Msun$. Integrating
over the accretion rate in Equation~\ref{eq_mdot}, we find that this
portion of the disk would be accreted after $t_{\rm acc} \sim
P_m$. For the parameters considered here, the orbital period of the
debris deepest in the SMBH potential well is $P_m \sim 2\times 10^8$s
and is shorter for more bound configurations of the envelope,
corresponding to RG orbits with lower initial eccentricity. This
implies that with a moderate or low level radiative feedback the gas
would be swallowed by the black hole on a time scale $\sim
10$yr. Alternatively, strong radiative feedback would promptly
disperse and ionize the debris. This time scale places a lower limit
constraint on the time between the disruption of the hydrogen envelope
and disruption of the helium core. Note that PS1-10jh was not detected
in 2009 in the deep coadd of the {\it GALEX} time domain survey with a
$\sim3\sigma$ upper limit of $>25.6$mag
\citep{gezari12}. In the context of our model this indicates that by
that point most of the disk must have been accreted or dispersed.

Another possible explanation for the weakness or absence of the
hydrogen emission lines is that the debris hydrogen envelope may be
present at the disruption site in the vicinity of the SMBH, but is
completely ionized or produces a negligible amount of hydrogen
emission. Tidal disruptions have indeed been proposed to create a
highly ionized nebular emission from the ISM in the inner kpc of of
their host galaxies \citep{eracleous95}. This scenario is considered
by \citet{guill13}, where 3d hydrodynamic simulations show the
disruption of a low mass main sequence star ($\sim 0.2\Msun$) with
debris forming a finite size BLR, truncated at the
outer radius. The compact BLR consists of gas from the disrupted star
including hydrogen, but effectively behaves as an HII region and
consequently emits weak hydrogen emission lines below the current
detection threshold. He~II emission-line on the other hand
remains present in the spectrum longer due to the higher ionization
potential of this element.  In this scenario, the unbound tidal
stream has a negligible surface area and makes negligible contribution
to either the continuum or line emission. This explanation is
consistent with the finding that the luminosity of the hydrogen Balmer
lines from the tidal disruption of a solar type star by a
$10^6\,\Msun$ SMBH is $\lesssim10^{39}{\rm erg\,s^{-1}}$
\citep{bogdanovic04}, and thus, below the detection threshold of the
available spectroscopic data for PS1-10jh. This work however does
not address the intensity of the helium emission-lines and hence does
not constrain the He/H line ratio.

Recently \citet{gaskell13} carried out the one-dimensional
photoionization modeling of the emission properties of gas in
truncated BLRs\footnote{In this approach, the photoionization
properties of the gas are calculated as a function of distance from
the ionization source.} and concluded that strong helium emission does
not require depletion of hydrogen. Specifically, they find that the
HeII $\lambda$4686/H$\alpha$ line ratio peaks at the value as high as
$\sim4.0$ when the gas density is $\sim10^{11}\,{\rm cm^{-3}}$, given
a solar abundance ratio of the two elements. This line ratio is a
sensitive function of the gas density however and the work by
\citet{gaskell13} also indicates that unless most of the solar
metallicity debris resides in the density range $10^{10} < n <
10^{12}\,{\rm cm^{-3}}$, the value of the line ratio HeII
$\lambda$4686/H$\alpha \lesssim 1$. Stellar debris produced in tidal
disruptions is indeed characterized by a wide range of densities and
is spatially distributed over eccentric orbits, allowing multiple
density phases of gas to reside at the same distance from the
ionization source \citep{bogdanovic04,strubbe09,strubbe11}. The
non-axisymmetric distribution of the gas results in a more complex
emission stratification than in the BLR regions of AGN that does not a
priory preclude the emission of the broad hydrogen lines. Indeed, more
recent and detailed calculations indicate that it is difficult to
suppress them to the level consistent with the observed properties of
PS1-10jh (L.E. Strubbe \& N. Murray 2014, in preparation).

Given a difficulty to "hide" the hydrogen emission lines in the
presence of the illuminated hydrogen rich stellar debris, in this work
we consider a family of models that lead to a near complete {\it
accretion} rather than {\it dispersal}, of the hydrogen envelope of
the partially disrupted red giant star.  In the context of this
hypothesis, at a later point in time when the helium core is disrupted,
helium dominates the optical emission line spectrum, as predicted by
the photoionization calculations of the spectrum from a tidally
disrupted white dwarf by \citet{sesana08}. This is also consistent
with the work \citet{clausen11} who find that the UV and optical
spectrum of a tidally disrupted carbon-oxygen white dwarf is dominated
by these species and characterized by the overall lack of hydrogen
emission lines. It is also worth noting that have the multiple and
more varied emission line features been seen in the spectrum of
PS1-10jh, it would be possible to put stronger constrains on the
structure of the disrupted red giant star and the mass of the black
hole. This is illustrated by \cite{clausen12} who found a good
agreement of their modeled emission line spectrum from the disruption
of a horizontal branch star by a BH with that observed from the
globular cluster NGC 1399 which hosts the ultraluminous X-ray source
CXOJ033831.8-352604.

\subsection{Distribution of the Helium-rich Debris on the Sky}

The favored portion of the parameter space, $ 6 < \beta < 24$ marked
in Figure~\ref{fig1}, maps into pericentric radii in the range $\sim
13-52\,r_g$.  At this distance the helium core and its debris are
subject to relativistic effects such as pericenter precession 
\citep[see][for discussion of this effect]{guill13} and
precession of the orbital plane, if the orbital axis and spin of the
SMBH are misaligned.

Evidence that relativistic effects can lead to the extended
distribution of the debris can be found in work by \citet{haas12} who
simulated the disruption of a white dwarf star by a spinning IMBH with
arbitrarily oriented spin axes. They find that as a consequence of
frame-dragging, the stellar debris forms an optically and
geometrically thick torus rather than an accretion disk. This leads to
the obscuration of the inner fallback disk by the outflowing
debris. As a consequence, the spectrum from the debris is softer
because it is mostly emitted by the material further away from the
black hole. Along similar lines \citet{lu97} have calculated the
temperature of radiation reprocessed and re-emitted at the Eddington
limit by an optically thick, static shell of gas with mass $\sim
0.1\Msun$ enshrouding the inner debris accretion disk
\begin{equation}
T_{\rm shell} \approx \left(\frac{L_E}{4\pi R_{\rm out}^2 \sigma}\right)^{1/4} = 
2.5\times10^4\,{\rm K}\,\left(\frac{M}{10^7 M_{\rm env}} \right)^{1/4},
\label{eq_Tshell} 
\end{equation}
where $R_{\rm out}$ is defined as the photospheric radius -- the
radius at which the optical depth to Thomson scattering $\tau_T =
1$. $T_{\rm shell}$ calculated in Equation~\ref{eq_Tshell} is
comparable to the characteristic temperature of radiation maintained
by PS1-10jh throughout the disruption. This finding, combined with the
fact that the range for the peak bolometric luminosity of PS1-10jh
inferred from observations permits values lower than the Eddington
luminosity \citep{gezari12}, leaves room for relativistic effects as a
plausible explanation for the configuration and emission properties of
the debris in PS1-10jh.

\section{Conclusions}\label{S_conclusions}

We consider the tidal interaction of a red giant star with a SMBH in
which the disruption is partial, leading to the tidal stripping of the
star's hydrogen envelope and subsequent inspiral of the compact helium
core toward the black hole.  Depending on its structure and the SMBH
mass, the helium core could either inspiral until it falls into the
SMBH or be tidally disrupted, giving rise to a second disruption flare
following the disruption of the hydrogen envelope.

During the phase in which the RG envelope is stripped, the orbital
evolution of the remnant is determined by tidal interaction which
dominates over the emission of GWs. The details of this process are
uncertain, cannot be evaluated analytically, and require 3d
hydrodynamic simulations to determine the dynamics of the compact
core. This deserves further attention because some fraction of the
remnant cores will evolve to become EMRIs detectable by the future
space-based GW observatories. The envelope effects operating on these
types of EMRIs can drive the frequency evolution of their GWs and
consequently determine how quickly they "migrate" into or out of the
frequency band of the detector, thus affecting their observability and
potentially providing a key feature for their identification.

Once the envelope is lost a pure helium core is not easily flexed by
the SMBH's tidal field and from that point on the emission of
gravitational radiation dominates the orbital evolution and tidal
dissipation plays a secondary role. Since the Kelvin-Helmoltz time
scale of helium white dwarfs is long ($\sim 10^9$yr), a significant
fraction of the dissipated energy can be retained within the RG
remnant. Whether or not perpetual tidal heating episodes lead to the
disruption of the remnant depends on its structure, as well as the
initial orbit and the SMBH mass. For example, we find that for compact
remnants with $M_{\rm core} = 0.3\Msun$ and $R_{\rm core} = 10^9{\rm
cm}$ the deposited energy is only a small fraction of their binding
energy and such surviving cores are possible progenitors of EMRIs.  On
the other hand, less massive and compact cores orbiting around $\sim
10^6\Msun$ SMBHs can be destroyed by tidal heating and hence, they
form a parent population for tidally disrupted cores.

In the case of a recently discovered tidal disruption candidate
PS1-10jh, we find that there is a set of orbital solutions at high
eccentricities, consistent with a nearly parabolic initial orbit of
the RG star, which lead to the accretion of the tidally stripped
hydrogen envelope by the SMBH before the inspiralling helium core is
disrupted. The allowed solutions confine the remnant to a portion of
the parameter space where tidal heating is intense and can result in
disruption of the core before it reaches the tidal radius.  In this
scenario, the hydrogen envelope ends up bound to the SMBH in its
entirety and is accreted on the time scale of about $10$ yr,
placing a lower limit on the time between the disruption of the
hydrogen envelope and disruption of the helium core.  On the other
hand, the helium debris produced by the subsequent disruption of the
RG core on a highly eccentric orbit is only partially bound, thus
giving rise to the canonical power-law decay of the late luminosity
curve observed in PS1-10jh. This sequence of events can provide one
plausible explanation for the puzzling absence of the hydrogen
emission lines from the spectrum of PS1-10jh, interpreted as an
apparent absence of the hydrogen envelope at the point in time when
the core is disrupted.

Furthermore, the indication that the UV and soft X-ray radiation
emitted from the nuclear debris disk is reprocessed by an intervening
shell of helium debris can be explained if one portion of the debris
assumes a geometrically extended, torus-like geometry around the black
hole. In the case of PS1-10jh the leading culprit responsible for this
configuration of the debris are radiation forces but also the
relativistic frame dragging due to the spinning SMBH.  If so, it is
intriguing to consider whether the light curve and spectrum of
PS1-10jh may contain any other imprints of the SMBH spin. In order to
provide an answer to that question, we need a more systematic
understanding of the dependance of the disruption signatures on the
black hole spin, which can be achieved through simulations of these
events.

The wide-field and high-cadence transient surveys such as {\it GALEX},
Pan-STARRS, Palomar Transient Factory, and the Large Synoptic Survey
Telescope, are expected to detect many more tidal disruption events
over the next decade. Therefore, the development of a new generation
of theoretical models is timely, as increasingly diverse tidal
disruption events are being captured in observations.

\acknowledgments
T.B. thanks Michael Eracleous, Suvi Gezari, Pablo Laguna,
Linda Strubbe, and Fabio Antonini for insightful comments and Lars
Bildsten for useful discussion about the properties of the ELM
WDs. T.B. acknowledges the support from the Alfred P. Sloan Foundation
under grant No. BR2013-016. P.A.S. acknowledges support from the
Transregio 7 ``Gravitational Wave Astronomy'' financed by the Deutsche
Forschungsgemeinschaft DFG (German Research Foundation).  This
research was supported in part by the National Science Foundation
under grant No. NSF PHY-1125915 and NSF AST-1333360.  The authors
acknowledge the hospitality of the Kavli Institute for Theoretical
Physics where one part of this work has been completed.



\end{document}